\documentclass[twocolumn,twocolappendix]{aastex631}
\usepackage{epsfig}
\usepackage[caption=false]{subfig}
\usepackage[]{xcolor}
\usepackage{amsmath}

\begin{document}

\title{No evidence that the binary black hole mass distribution evolves with redshift}

\author[0000-0002-2254-010X]{Max Lalleman}
\affiliation{Universiteit Antwerpen,
Prinsstraat 13,
2000 Antwerpen, België}
\email{max.lalleman@uantwerpen.be}

\author[0000-0002-9296-8603]{Kevin Turbang}
\affiliation{Universiteit Antwerpen,
Prinsstraat 13,
2000 Antwerpen, België}
\affiliation{Theoretische Natuurkunde, Vrije Universiteit Brussel, Pleinlaan 2, B-1050 Brussels, Belgium}

\author[0000-0001-9892-177X]{Thomas Callister}
\affiliation{Kavli Institute for Cosmological Physics, The University of Chicago, 5640 S. Ellis Ave., Chicago, IL 60615, USA}

\author[0000-0003-4180-8199]{Nick van Remortel}
\affiliation{Universiteit Antwerpen,
Prinsstraat 13,
2000 Antwerpen, België}

\begin{abstract}
The mass distribution of merging binary black holes is generically predicted to evolve with redshift, reflecting systematic changes in their astrophysical environment, stellar progenitors, and/or dominant formation channels over cosmic time.
Whether or not such an effect is observed in gravitational-wave data, however, remains an open question, with some contradictory results present in the literature.
In this paper, we study the ensemble of binary black holes within the latest GWTC-3 catalog released by the LIGO-Virgo-KAGRA Collaboration, systematically surveying for possible evolution of their mass distribution with redshift.
We specifically focus on two key features present in the binary black hole primary mass distribution -- (1) an excess of $35\,M_\odot$ black holes and (2) a broad power-law continuum ranging from 10 to $\gtrsim 80 M_\odot$ -- and ask if one or both of these features are observed to vary with redshift.
We find no evidence that either the Gaussian peak or power-law continuum components of the mass distribution change with redshift.
In some cases, we place somewhat stringent bounds on the degree of allowed redshift evolution.
Most notably, we find that the mean location of the $35\,M_\odot$ peak and the slope of the power-law continuum are constrained to remain approximately constant below redshift $z\approx 1$.
The data remain more agnostic about other forms of redshift dependence, such as evolution in the height of the $35\,M_\odot$ excess or the minimum and maximum black hole masses.
In all cases, we conclude that a redshift-dependent mass spectrum remains possible, but that it is not required by current data.
\end{abstract}

\section{Introduction}\label{sec:intro}
To date, the LIGO-Virgo-KAGRA Collaboration \citep[LVK,][]{aasi_advanced_2015,acernese_advanced_2015,akutsu_overview_2021} has published the detection of ninety gravitational-wave signals from merging compact binaries~\citep{PhysRevX.13.041039}.
This growing body of gravitational-wave observations is providing ever more detailed information about the properties of compact binary mergers~\citep{PhysRevX.13.011048,Callister:2024cdx}, offering a census of their masses, spins, and the merger rate in the local Universe.

The majority of observed gravitational waves originates from the mergers of stellar-mass binary black holes in the relatively local Universe.
Gravitational-wave observations are not limited to the local Universe, however; the detection horizon of the Advanced LIGO and Virgo network now extends to or beyond redshift $z\approx 1$~\citep{PhysRevX.13.041039,capote_advanced_2024}.
In addition to studying the demographics of local compact binary mergers, we therefore have an opportunity to study how these demographics systematically evolve over cosmic time.
The merger rate of binary black holes, for example, is observed to increase with redshift \citep{PhysRevX.13.011048,callister2023parameterfree,edelman_cover_2023}, and binary black holes merging at earlier cosmic times may have had different spin distributions than those merging today~\citep{Biscoveanu_2022,heinzel_probing_2024}.

Within this context, a commonly asked question is whether the mass distribution of merging black holes evolves with redshift.
A redshift-dependent mass distribution is a generic and robust astrophysical prediction, arising from a variety of effects in a variety of different astrophysical scenarios:

\textit{Isolated binaries.}
The efficiency of massive black hole formation and merger is expected to depend sensitively on the metallicities of progenitor stars~\citep{Belczynski_2010, Belczynski_2016, 10.1093/mnras/sty3087, 10.1093/mnras/stz1150, 2020ApJ...898..152S,2024arXiv241102484V}.
The overall evolving chemical enrichment of the Universe may therefore produce systematic shifts in the masses of merging black holes, with more massive black holes preferentially arising from low-metallicity stars prevalent at larger redshifts.
In the extreme limit, Population III stars formed at high redshifts from pristine primordial gas may even avoid pair-instability~\citep{Woosley_2021}, collapsing to yield massive black holes falling in or above the pair instability ``mass gap'' avoided by subsequent generations of exhibited~\citep{Liu_2020,Tanikawa_2021} (although see~\cite{Costa_2023}). 
Among binaries formed in isolation, correlations between masses and evolutionary delay times may also impart redshift dependence to the black hole mass spectrum.
Low-mass binaries, for example, may be more prone to unstable mass transfer leading to a common envelope, merging more rapidly than high-mass binaries evolving via stable mass transfer and causing an observed shift towards \textit{lighter} black holes with higher redshift~\citep{van_Son_2022}.

\textit{Mergers in dense clusters.}
Black holes merging in globular clusters, young star clusters, and/or nuclear clusters~\citep[e.g.,][]{portegies_zwart_black_2000,2010MNRAS.407.1946D,2011MNRAS.416..133D,2015PhRvL.115e1101R}, although dynamically influenced by many-body encounters, are also affected by metallicity-dependent stellar evolution, and may therefore exhibit a redshift-dependent mass distribution~\citep{10.1093/mnras/stab1334,10.1093/mnras/stac422,torniamenti2024hierarchicalbinaryblackhole}.
Dense clusters can also foster the assembly of massive black holes via repeated ``hierarchical" mergers~\citep{Antonini_2016, Fishbach_2017, Antonini_2019, Gerosa_2021, Zevin_2022, mahapatra2024predictionssimpleparametricmodel}.
These hierarchical mergers preferentially occur early in the lifetime of clusters, such that the masses of merging black holes are again systematically larger at large redshifts~\citep{Antonini_2016,Ye_2024,torniamenti2024hierarchicalbinaryblackhole}.

\textit{Active galactic nuclei.}
Black hole mergers may instead be produced in the accretion disks of active galactic nuclei~\citep[AGN;][]{mckernan_intermediate_2012,2016ApJ...819L..17B,2017MNRAS.464..946S,10.1093/mnras/stac2861}.
Merger products may themselves become trapped in the accretion disk, leading to possibly large numbers of hierarchical mergers that build up ever more massive black holes.
Massive hierarchical mergers become increasingly prevalent late in an AGN's lifetime~\citep{2024arXiv241018815D}, possibly yielding a redshift-dependent mass distribution.

\textit{Multiple formation channels.}
Finally, evolution of the black hole mass function with redshift may arise simply from the presence of multiple binary formation channels.
Different formation channels generically predict distinct black hole mass distributions and naturally vary in prevalence as a function of redshift.
As mixture fractions between formation channels evolve, so too will the overall mass distribution~\citep[e.g.,][]{Antonini_2016,Zevin_2021,2021PhRvD.103h3021W,10.1093/mnras/stac422,torniamenti2024hierarchicalbinaryblackhole,2024arXiv241108658F}.

Some authors, \cite{Farr_2019} and \cite{Mastrogiovanni_2021}, have also discussed the possibility of measuring the Hubble expansion $H_0$ using gravitational-wave sirens assuming a certain mass-redshift dependence of the binary black hole mass distribution, may that be redshift dependent or independent.
The precise shape and redshift dependence of the mass distribution is important since it helps breaking the degeneracy between mass and redshift of the sources and influences the inference of cosmological parameters.

Motivated by these predictions, several studies have searched for a redshift evolution in the observed black hole mass spectrum, with sometimes conflicting results.
\cite{Fishbach_2021} searched for redshift dependence in the maximum observed black hole mass, using binary black holes from the GWTC-2 catalog~\citep{PhysRevX.11.021053}.
Meanwhile, ~\cite{van_Son_2022} investigated relative prevalence of stable mass transfer vs. common envelope evolution among isolated binaries, searching GWTC-2 data for a resulting redshift-dependent mass distribution among binary black holes~\citep{PhysRevX.11.021053}.
Both studies yielded null results.
The study performed in~\cite{Fishbach_2021} was later repeated by the LVK Collaboration using additional data from GWTC-3~\citep{PhysRevX.13.011048}, with a redshift dependence remaining unobserved.
\cite{2023ApJ...957...37R} and~\cite{2024arXiv240616844H} instead employed flexible ``non-parametric'' methods, based on the discretization of the population into a large number of piecewise constant bins, to measure the black hole mass distribution across a range of redshifts.
Although this increased model flexibility translated into large uncertainties on the high-redshift mass distribution, neither study found a correlation between black hole masses and redshifts.

On the other hand, \cite{Karathanasis_2023} instead argued that current gravitational-wave data favor a large systematic shift towards higher masses between $z=0$ and $z=1$, a result they interpret as metallicity dependence in the pair-instability supernovae limit for black hole masses~\citep{2002ApJ...567..532H,farmer_mind_2019}.
More recently,~\cite{Rinaldi} obtained similar conclusions, identifying an even more dramatic evolution of the black hole mass distribution with redshift, such that the most common black hole masses shift from $\sim$ 10 solar masses in the local Universe to $\sim$ 50 M$_\odot$ at $z$ = 1.

Our goal in this paper is to revisit the question of whether current gravitational-wave data point to a redshift-dependent black hole mass distribution, adopting a more flexible and comprehensive approach in order to survey and constrain (or identify) a variety of possible forms such evolution could take.
Gravitational-wave data confidently support the existence of two features in the black hole mass spectrum: an excess of mergers with primary masses $m_1 \approx 35\,M_\odot$, and a broad power-law continuum ranging between 10 to 80 $M_\odot$~\citep{Abbott_2021, tiwari_exploring_2022,PhysRevX.13.011048,edelman_cover_2023,farah_things_2023,callister2023parameterfree}.
We focus on each of the features in turn, asking if either or both exhibit any redshift dependence in their parameters.
We find that current data show no evidence for a redshift-dependent binary black hole mass distribution and in some cases place informative limits on the degree of allowed evolution.
However, although no evidence for redshift evolution is found, we note that current results do not exclude a redshift dependence either:
future analyses with expanded gravitational-wave catalogs could reveal (or further limit) evidence for a cosmically varying black hole mass distribution.

The remainder of this paper is organized as follows.
In Section \ref{sec:method} we describe our analysis, include the precise models we use to search for and constrain a redshift-dependent mass spectrum.
In Section \ref{sec:results}, we present and discuss our resulting constraints for the redshift dependence of the mass distribution, while in Section \ref{sec: reverse sector}, we discuss the merger rate mass-dependence.
Finally, we conclude and discuss the astrophysical implications of this work in Section \ref{sec:conclusion}.

\section{Methods and data}
\label{sec:method}

\begin{figure*}
    \centering
    \includegraphics[width=0.48\textwidth]{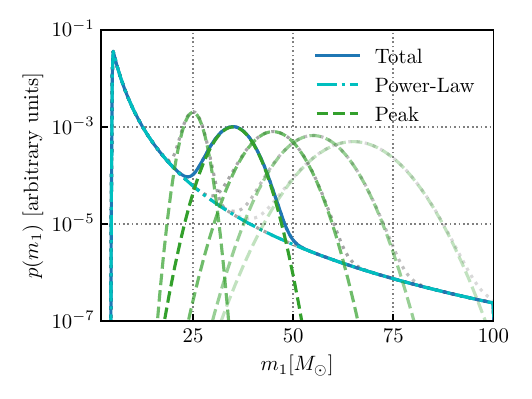}
    \includegraphics[width=0.48\textwidth]{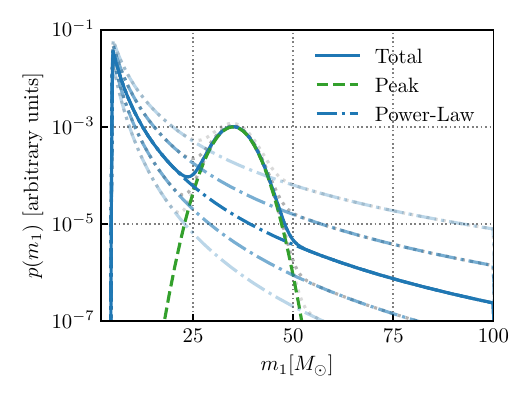}
    \caption{
    Cartoon illustrating the possible manners, as considered in this work, in which the binary black hole primary mass distribution might evolve with redshift.
    As illustrated in the left-hand panel, we consider the possibilities that the location, width, and/or height of the $35\,M_\odot$ excess evolve with redshift (see Sec.~\ref{sec:peak-sector}.
    And as illustrated on the right, we allow for evolution of the height, slope, and endpoints of the power-law continuum (see Sec.~\ref{sec:power-law-sector}).}
    \label{fig:cartoon-pl}
\end{figure*}

The primary mass distribution of merging binary black holes is well-characterized by a superposition of a Gaussian peak near $m_1 \approx 35\,M_\odot$ and a broad power-law distribution $p(m_1)\propto m_1^\alpha$, with $\alpha\approx -3.5$.
We therefore adopt the following parameterization as a baseline model for the black hole primary mass distribution~\citep{talbot_measuring_2018,Abbott_2021}:
\begin{widetext}
\begin{equation}
\label{eq:pm1}
    p(m_1) \propto \left\{
    \begin{array}{ll}
        \mathcal{A}\,(1 - f_{\rm p})\,m_1^{\alpha} + \mathcal{B}\,f_{\rm p}\,e^{\frac{-\left(m_1 -  \mu_{m}\right)^2}{2 \sigma_{m}^2}} & \mbox{($M_{\rm min}$ $\leq$ $m_1$ $\leq$ $M_{\rm max}$)} \\
        0 & \mbox{otherwise}.
    \end{array}
\right.
\end{equation}
\end{widetext}
Here, $\alpha$ is the slope of the power-law component and $\mu_m$ and $\sigma_m$ are the mean and standard deviation, respectively, of the Gaussian component.
The hyperparameter $f_{\rm p}$ governs the relative height of each component, while $\mathcal{A}$ and $\mathcal{B}$ are normalization constants.
The goal of this work is to investigate a possible redshift dependence in Eq. (\ref{eq:pm1}).
Concretely, we mainly focus on two questions:
    \begin{enumerate}
        \item Do the inferred hyperparameters of the Gaussian excess -- its mean, variance, and height -- vary with redshift?
        \item Do the inferred hyperparameters of the power-law continuum -- its boundaries, its slope, and/or the fraction of events in the continuum -- vary with redshift?
    \end{enumerate}
Both of these scenarios are illustrated schematically in Figure \ref{fig:cartoon-pl}, which shows a varying Gaussian peak (left) and a varying power-law continuum (right).

\subsection{Modeling a redshift-dependent mass distribution}\label{sec: varying information}

In order to answer the above questions, we promote the hyperparameters in Eq. (\ref{eq:pm1}) to functions of redshift.
In particular, we allow the hyperparameters of interest to vary as sigmoids, enabling a flexible and smooth transition between low- and high-redshift values.
In particular, if we let $\Lambda$ signify an arbitrary hyperparameter (e.g. $\alpha$, $\mu_m$, etc.), then we assume that $\Lambda$ evolves as
\begin{equation}\label{eq: sigmoid delta}
    \Lambda(z) = \frac{\Lambda_{\rm high} - \Lambda_{\rm low}}{1 + \exp\left[-\frac{1}{\Delta z_\Lambda}\left(z - \bar z_\Lambda\right) \right]} + \Lambda_\mathrm{low}.
\end{equation}
where $\Lambda_{\mathrm{low}}$ is the hyperparameter value at redshift $z\ll \bar z_\Lambda$ and $\Lambda_{\mathrm{high}}$ is the value approached asymptotically at $z \gg \bar z_\Lambda$.
The transition occurs across a redshift interval of a width $\Delta z_\Lambda$ centered at $\bar z_\Lambda$.

\subsection{Defining the differential merger rate}

Alongside the (redshift-dependent) primary mass distribution, we simultaneously measure the binary black hole secondary mass distribution, the spin distribution, and the overall evolution of the merger rate with redshift.
Our full model for the source-frame merger rate of binary black holes takes the form
\begin{equation}\label{eq:full merger rate}
\begin{aligned}
&\mathcal{R}(m_1,m_2,\vec\chi_1,\vec\chi_2;z) \\
    &\quad = \frac{dN}{dV_c\,dt_s\,dm_1 dm_2 d\chi_1 d\chi_2}(m_1,m_2,\vec\chi_1,\vec\chi_2;z) \\
    &\quad = R_{\rm ref}
        \frac{f(z)}{f(0.2)}
        \frac{\phi(m_1|z)}{\phi(20\,M_\odot|0.2)}
        p(m_2) p(\vec \chi_1) p(\vec \chi_2).
\end{aligned}
\end{equation}
The above quantity denotes the number of mergers per unit comoving volume $dV_c$, per unit source frame time $dt_s$, per unit source parameters.
The function $\phi(m_1|z)$ encodes the primary mass distribution at a given redshift.
This is given by Eq. (\ref{eq:pm1}), with additional smoothing factors that truncates the mass distribution at sufficiently low and high values $M_{\rm min}$ and $M_{\rm max}$:
\begin{equation}
\label{eq:phi_m}
    \phi\left(m_1\right|z)=\begin{cases}
    p\left(m_1|z\right) \exp \left(\frac{-\left(m_1-M_{\rm min}(z)\right)^2}{2 \delta m_{\rm {min}}^2(z)}\right),
        & \left(m_1 < M_{\rm min}(z)\right) \\
    p\left(m_1|z\right) \exp \left(\frac{-\left(m_1-M_{\rm max}(z)\right)^2}{2 \delta m_{\rm max}^2(z)}\right),
        & \left(m_1 > M_{\rm max}(z)\right) \\
    p\left(m_1|z\right) 
        & \left(\mathrm{else}\right)
\end{cases}.
\end{equation}
Here, $p(m_1 | z)$ is as given Eq. (\ref{eq:pm1}), but with hyperparameters promoted to functions of redshift as described in Section \ref{sec: varying information}.
Similarly, the hyperparameters controlling the truncation of the mass distribution, like $M_{\mathrm{min}}$ and $M_{\mathrm{max}}$ and the widths of the smoothening exponentials $\delta m_{\mathrm{min}}$ and $\delta m_{\mathrm{max}}$, are also regarded as functions of redshift using the aforementioned sigmoid formalism.

Within Eq.~\eqref{eq:full merger rate}, the function $f(z)$ captures the overall (mass-independent) evolution of the merger rate with redshift.
If merging black holes are of stellar origin, then the merger rate is likely to qualitatively trace cosmic star formation, initially rising as a function of redshift before peaking and decreasing at higher redshift~\citep{annurev:/content/journals/10.1146/annurev-astro-081811-125615, Madau_2017}.
Accordingly, we adopt a model~\citep{fishbach_does_2018,Callister_2020},
\begin{equation}
\label{eq:f-z}
f(z)=\frac{(1+z)^{\alpha_z}}{1+\left(\frac{1+z}{1+z_p}\right)^{\alpha_z+\beta_z}},
\end{equation}
which grows as $f(z) \propto (1+z)^{\alpha_z}$ at $z\ll z_p$ and decreases as $f(z) \propto (1+z)^{-\beta_z}$ at $z \gg z_p$.

We also assume a power-law distribution for the secondary mass of the binary, such that~\citep{fishbach_picky_2020}
\begin{equation}\label{eq: p(m_2)}
    p(m_2| m_1, \beta_q) \propto \frac{1 + \beta_q}{m_1^{1+\beta_q} - M_{\rm min}^{1+\beta_q}} m_2^{\beta_q}.
\end{equation}
In principle, we could additionally include the power-law slope $\beta_q$ among the parameters we allow to vary with redshift.
When doing so, however, only uninformative constraints, and so for simplicity exclude variation of $\beta_q$ in the analyses discussed in Section \ref{sec:results}.
Finally, our models for the probability distributions of component spins $\vec \chi_1$ and $\vec\chi_2$ are described in Appendix~\ref{app: prior}.

\subsection{Data and likelihood}

In this work, we consider all binary black hole events in the LIGO-Virgo-KAGRA Collaboration's GWTC-3 catalog~\citep{PhysRevX.13.041039} with false alarm rates below one per year.

Given a binary black hole population described by hyperparameters $\Lambda$, the likelihood of an observed gravitational-wave catalog is ~\citep{10.1063/1.1835214, PhysRevD.98.083017, 10.1093/mnras/stz896,vitale_inferring_2020}
\begin{equation}
\label{eq: p_BBH}
  p_{\mathrm{BBH}}\left(\left\{d_{i}\right\}|\Lambda\right)\propto e^{-N_{\rm exp}\left(\Lambda\right)} \prod_{i=1}^{N_{\mathrm{obs}}}\int p\left(d_{i}|\lambda\right) \frac{dN}{d\lambda}(\Lambda) d \lambda.
\end{equation}
Here, $N_{\rm obs}$ is the total number of gravitational-wave observations, with data represented by $\{d_i\}_{i=1}^{N_{\rm obs}}$, and $N_\mathrm{exp}(\Lambda)$ is the total number of expected mergers (observed or unobserved) predicted to occur over the given observation period.
In the above likelihood, $\lambda$ signifies the individual parameters (component masses, spins, redshift, etc.) of each binary; the likelihood of the $i$th gravitational-wave event given parameters $\lambda$ is denoted $p(d_i|\lambda)$.
The quantity $dN/d\lambda(\Lambda)$ is the detector-frame rate of gravitational-wave events, related to the source-frame rate in Eq.~\eqref{eq:full merger rate} by
\begin{equation}
\begin{aligned}
    \frac{dN}{d\lambda} &\equiv \frac{dN}{dm_1\,dm_2\,dz\,d\vec\chi_1\,d\vec\chi_2} \\
    &= \frac{dV_c}{dz}\frac{T_\mathrm{obs}}{1+z}\mathcal{R}(m_1,m_2,\vec\chi_1,\vec\chi_2;z),
\end{aligned}
\end{equation}
where the factor $(1+z)^{-1}$ transforms from detector-frame to source-frame time, $T_{\rm obs}$ is the observing time, and $dV_c/dz$ gives the comoving volume per unit redshift.

In practice, we evaluate Eq.~\eqref{eq: p_BBH} not as an integral over events' likelihoods, but instead via Monte Carlo averages over discrete posterior samples for each event.
Given sets of samples $\{\lambda\} \sim p(\lambda|d_i)$ drawn from each event's posterior, Eq.~\eqref{eq: p_BBH} may be approximated via
\begin{equation}
\label{eq: p_BBh sampled}
    p_{\rm BBH}\left(\left\{d_{i}\right\}|\Lambda\right)\propto e^{-N_{\rm exp}\left(\Lambda\right)} \prod_{i=1}^{N_{\rm obs}}\left\langle\frac{dN/d\lambda(\lambda_i)}{p_{\rm pe}(\lambda_{i})}\right\rangle_{\text {samples}},
\end{equation}
where $p_{\rm pe}(\lambda_i)$ is the prior utilized during parameter estimation and $\langle \cdot \rangle_\mathrm{samples}$ indicates an average taken over posterior samples of a given event.

Search selection effects are captured by the expected number of detections, $N_\mathrm{exp}(\Lambda)$, given by
\begin{equation}
\label{eq: total expected number of merger evens}
    N_{\rm exp}(\Lambda)=\int d \lambda P_{\text{det}}(\lambda) \frac{dN}{d\lambda}(\lambda|\Lambda),
\end{equation}
where $P_{\mathrm{det}}(\lambda)$ is the probability of detecting an event with event parameters $\lambda$.
This integral, too, can be approximated as a Monte Carlo average over simulated signals injected into and recovered from gravitational-wave data.
Given a total number $N_\mathrm{inj}$ of such injections, each drawn from a parent distribution $p_\mathrm{inj}(\lambda)$,
\begin{equation}
    \label{eq:MC_selection_effects}
    N_{\rm exp}(\Lambda) \approx \frac{1}{N_{\rm inj}}\sum_i^{N_{\rm found}}\frac{dN/d\lambda_i}{p_{\rm inj}(\lambda_i)},
\end{equation}
where the summation runs over the subset of successfully recovered injections.

More information about the exact data used in our analysis can be found in Appendix~\ref{app: data}.

\section{Does the black hole mass spectrum evolve with redshift?} \label{sec:results}

As discussed previously, the binary black hole primary mass distribution can be phenomenologically modeled as two pieces: (\textit{i}) a Gaussian excess situated on top of (\textit{ii}) a broad power-law continuum.
We search for redshift evolution within each of these features in turn.
In Section ~\ref{sec:peak-sector}, we first explore the degree to which the Gaussian peak evolves with redshift.
Then, in Section ~\ref{sec:power-law-sector}, we examine possible redshift evolution in the power-law continuum.\footnote{Note that, because $f_p$ controls the prominence of both the power-law and Gaussian components of the primary mass distribution, the analyses in Secs.~\ref{sec:peak-sector} and \ref{sec:power-law-sector} are not strictly independent of one another.}
In Section ~\ref{sec: reverse sector} we reverse our perspective, and instead analyze GWTC-3 data in search of mass dependence in the redshift distribution of binary black holes.

\subsection{Does the Gaussian excess vary with redshift?}
\label{sec:peak-sector}

The Gaussian excess observed in the black hole primary mass distribution is governed by three parameters: the mean of the peak $\mu_m$, its standard deviation $\sigma_m$, and the mixing fraction $f_\mathrm{p}$ controlling its relative height.
We allow each of these hyperparameters to vary as a function of redshift, as in Eq. \eqref{eq: sigmoid delta}, using the priors described in Appendix \ref{app: prior}.

Figure \ref{fig:peak-R-m} shows our resulting constraints on the redshift-dependent mass distribution when the Gaussian peak is allowed to evolve.
In particular, we show the probability distribution of primary masses taken at two different redshifts: $z = 0.2$ (upper subplot) and $z = 0.75$ (lower subplot).
Green lines correspond to individual draws from our population posterior, and solid black lines bound the $95\%$ credible intervals on $p(m_1|z)$.
For reference, the dashed black lines in the lower subplot show the $95\%$ credible constraints at $z=0.2$ above.

We see in Fig.~\ref{fig:peak-R-m} that the \textit{location} of the peak is constrained to remain reasonably stationary across the full range of redshifts considered.
At the same time, its height is less well-constrained; it remains \textit{possible} for the $35\,M_\odot$ peak to become more pronounced at higher redshift, indicating that a growing peak is not excluded by current observations.
This growth is not required, however, indicating that current gravitational wave catalogs contain no affirmative evidence for a redshift-dependent peak.

\begin{figure}
    \centering
    \includegraphics[width=0.45\textwidth]{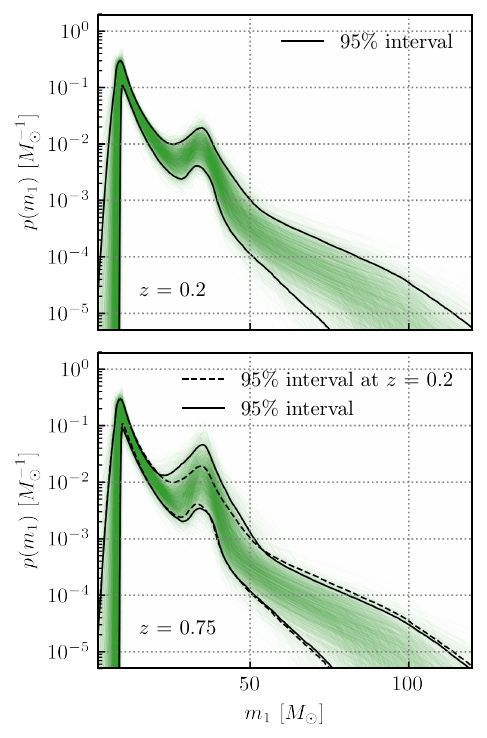}
    \caption{
    The inferred binary black hole primary mass distribution, when allowing the Gaussian excess at $\sim 35\,M_\odot$ to evolve with redshift.
    \textit{Upper panel:} The inferred primary mass distribution at $z=0.2$.
    Green traces illustrate individual draws from our hyperposterior, while solid black lines mark $95\%$ credible bounds.
    \textit{Lower panel:} The primary mass distribution inferred at $z=0.75$.
    For comparison, the dashed black lines illustrate the $95\%$ credible bounds from the upper subplot at $z=0.2$.
    At that higher redshift, we see elevated uncertainty in the primary mass distribution.
    However, we observe no systematic deviations between the distributions inferred at redshifts $z=0.2$ and $1$.}
    \label{fig:peak-R-m}
\end{figure}

\begin{figure}
    \centering
    \includegraphics[width=0.45\textwidth]{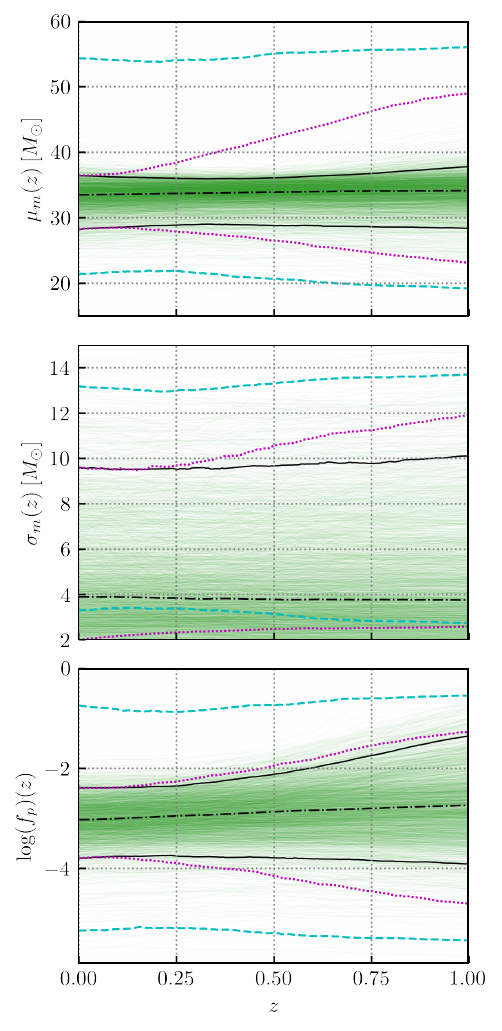}
    \caption{
    Inferred values of the hyperparameters characterizing the mean (\textit{top}), standard deviation (\textit{middle}), and height (\textit{bottom}) of the Gaussian peak in the black hole mass spectrum, as a function of redshift.
    Within each subplot, green traces mark individual hyperposterior samples, while solid and dot-dashed black curves indicate $95\%$ credible posterior bounds and medians, respectively.
    Cyan dashed lines analogously illustrate $95\%$ credible prior bounds.
    Finally, the dotted magenta curves indicate $95\%$ credible prior bounds on each parameter when first \textit{conditioned} on measured posteriors at $z=0$.
    We see that location of the Gaussian peak is constrained to remain approximately constant over a large range of redshift, while evolution in the width of the Gaussian is nearly unconstrained.
    Finally, data disfavor a Gaussian excess that \textit{shrinks} with redshift, but are consistent with a peak that remains constant or grows towards larger redshifts.}
    \label{fig:sigmoid-peak}
\end{figure}

An alternative view of our results is given in Fig.~\ref{fig:sigmoid-peak}, which shows our posteriors on the relevant hyperparameters themselves as a function of redshift.
Specifically, the top, middle and lower panels show our posteriors on the location $\mu_m(z)$, width $\sigma_m(z)$, and logarithmic height of the peak $\log f_p(z)$.
Each individual green trace shows a single posterior draw, the black dot-dashed line shows the median posterior values as a function of redshift, and the solid black lines bound central 95$\%$ credible intervals on each hyperparameter.
For reference, the dashed cyan lines show the 95$\%$ intervals from our prior distributions on $\mu_m(z)$, $\sigma_m(z)$, and $\log f_p(z)$.

In all three cases (and for $\mu_m(z)$ and $\log f_p(z)$ in particular) our posteriors are constrained away from our priors across a broad range of redshifts.
At face value, however, the interpretation of these measurements is ambiguous:
Have we indeed meaningfully measured these parameters across a range of redshifts?
Or have we only succeeded in measuring them at $z\approx 0$, with the high-redshift posteriors constituting simply a prior-dependent extrapolation of these low-redshift bounds?

To answer this question, we show as dotted magenta curves the \textit{conditional priors} on each parameter, given our posteriors at $z=0$.
These curves illustrate the remaining redshift dependence allowed by our prior, when informed only by data at $z=0$, and accordingly demonstrate the degree to which low-redshift information is or is not being extrapolated to high redshifts.

Comparing our posterior on $\mu_m(z)$ to this conditional prior, we see that the posterior is constrained well away from the conditional prior at high redshifts.
Thus the conclusion that $\mu_m(z)$ must remain relatively constant is a feature of the data, not an artifact of our prior.
Our posterior on $\sigma_m(z)$, meanwhile, is marginally constrained away from the conditional prior, but it is apparent that additional data will be required before one can obtain informative constraints on the width of the $35\,M_\odot$ peak with redshift.
Our lower bound on $\log f_p(z)$, in turn, is strongly constrained away from the conditional prior, such that we can confidently conclude that the $35\,M_\odot$ peak \textit{does not shrink with redshift}.
In contrast, the upper bound on $\log f_p(z)$ coincides with the conditional prior, such that any apparent growth in the $35\,M_\odot$ peak (e.g. the lower panel of Fig.~\ref{fig:peak-R-m}) is purely a prior effect.
As shown by the conditional priors, the results at $z \geq$ 0.5 are not just extrapolations from $z$ = 0, but the results at $z \approx$ 1 are probably extrapolations from intermediate redshift.

A corner plot showing the full posterior on the hyperparameters governing $\mu_m(z)$ is given in Appendix~\ref{app: inference results}.  

\subsection{Does the power-law continuum evolve with redshift?}\label{sec:power-law-sector}

Next, we instead ask if observed binary black holes exhibit evidence for a redshift-dependent power-law continuum.
This continuum is defined by a set of six parameters:
a power-law index $\alpha$, the truncation points $M_\mathrm{min}$ and $M_\mathrm{max}$, the mass scales $\delta m_\mathrm{min}$ and $\delta m_\mathrm{max}$ over which the truncation occurs, and the relative height of the continuum given by $1-f_p$.

\begin{figure}
\centering
    \includegraphics{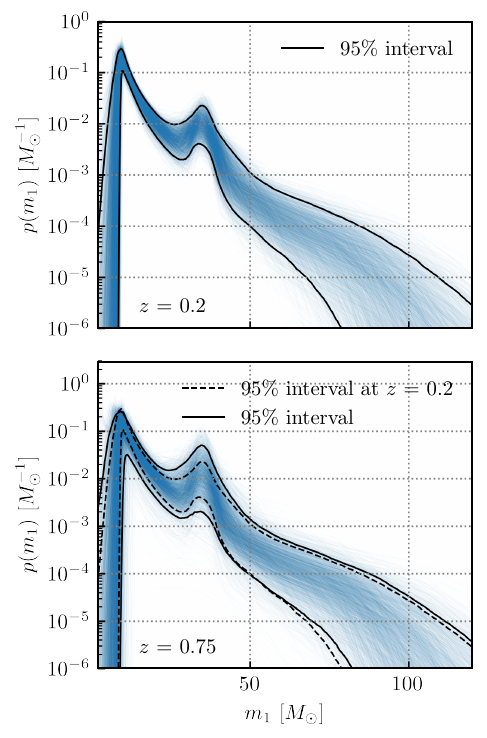}
    \caption{
    As in Fig.~\ref{fig:peak-R-m}, but now allowing the power-law continuum of the binary black hole primary mass spectrum to evolve with redshift.
    We again see that, although the primary mass distribution is more uncertain at large redshifts, the data indicate no systematic evolution between $z=0.2$ and $z=0.75$.
    }
\label{fig: merger-rate-for-different-z}
\end{figure}

\begin{figure}
\centering 
  \includegraphics{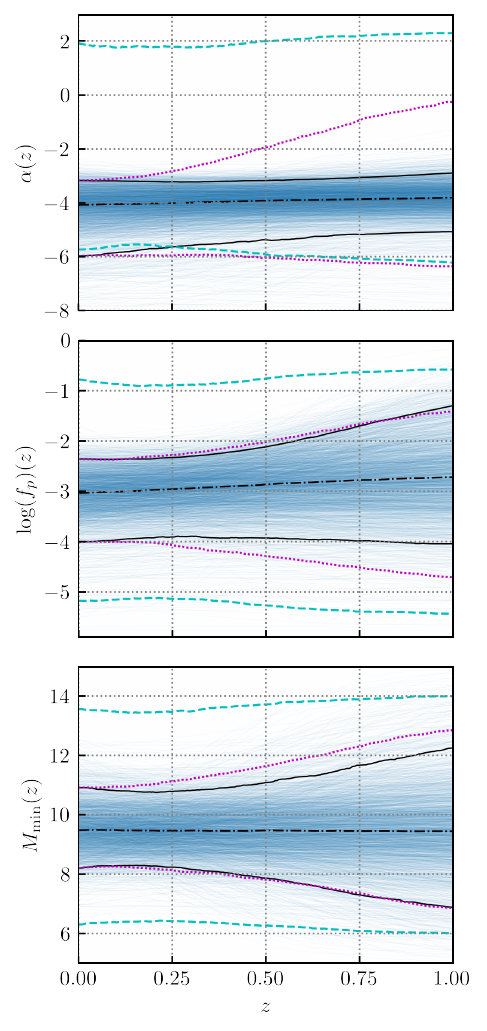}
  \caption{
    As in Fig.~\ref{fig:sigmoid-peak}, but now showing posteriors on the redshift-dependent hyperparameters characterizing the power-law continuum.
    Specifically, we show posteriors on the power-law slope (\textit{top}), mixing fraction (\textit{middle}), and the location below which the mass distribution is truncated (\textit{bottom}).
    Posteriors on other parameters (such as the maximum black hole mass and high/low mass truncation scales) are uninformative and not shown here.
    The most precisely measured parameter is the power-law slope, which is constrained to remain approximately constant below $z=1$.
  }
  \label{fig:traces-power-law}
\end{figure}

Figure~\ref{fig: merger-rate-for-different-z} shows the resulting constraints on the black hole primary mass distribution, at both $z=0.2$ (upper subplot) and $z=0.75$ (lower subplot).
Solid black lines trace 95\% credible posterior bounds at each redshift, while the dashed lines in the lower subplot give the $95\%$ credible bounds at $z=0.2$ for ease of comparison.
As in Fig.~\ref{fig:peak-R-m}, there is little indication that the black hole mass distribution evolves with redshift.
The only noticeable difference in $p(m_1)$ between high and low redshifts is the possibility of a change in the height of the power-law continuum (compensated by an equal and opposite change in the height of the Gaussian peak).
We cannot, of course, rule out redshift variation on scales finer than we can currently probe with existing data.

In Fig.~\ref{fig:traces-power-law}, we plot posteriors on relevant hyperparameters as a function of redshift.
As in Fig.~\ref{fig:sigmoid-peak} above, solid black lines indicate $95\%$ credible posterior bounds, dashed cyan lines mark the $95\%$ bounds imposed by our prior, and dotted magenta lines give the $95\%$ prior bounds once conditioned on our posterior at $z=0$.
We specifically show redshift-dependent constraints on the continuum power-law index $\alpha(z)$, the log mixing fraction, $\log f_\mathrm{p}(z)$, and the minimum mass $M_\mathrm{min}(z)$ below which the primary mass distribution is truncated.
Our posteriors on other parameters, such as the maximum black hole mass $M_\mathrm{max}$ and truncation scale lengths, do not deviate from their corresponding conditional priors and are therefore not shown.

The power-law index $\alpha(z)$ is well-measured across a range of redshifts, bounded confidently away (both above and below) from the conditional prior.
At the same time, we see no evidence for a systematic shift in this slope.
The posterior on the mixing fraction $f_p(z)$ is quite similar to behavior seen in Fig.~\ref{fig:sigmoid-peak}, bounded above the conditional prior given by the measured continuum and peak heights at $z=0$.
It therefore remains possible that the height of the power-law continuum decreases with redshift (offset by an increase in the height of the Gaussian peak), but the data show no affirmative evidence of this behavior.
Finally, we see minimal information regarding the evolution of $M_\mathrm{min}$ with redshift: its posterior shows a minimal shift relative to the conditional prior on $M_\mathrm{min}(z)$, but it remains possible for $M_\mathrm{min}$ to both rise or fall with redshift.

Corner plots showing the full posterior hyperparameters controlling $\log f_p(z)$ and $\alpha(z)$ are given in Appendix~\ref{app: inference results}.  

\section{Does the black hole redshift distribution vary with mass?}
\label{sec: reverse sector}

In Section~\ref{sec:results}, we explored constraints on the systematic evolution of the binary black hole primary mass distribution with redshift, finding no evidence for any evolution.
This is not the only way to frame this observational question, though.
The binary black hole merger rate is known to rise as a function of redshift.
Therefore, instead of asking if the black hole mass distribution depends on redshift, one might invert this question and ask: \textit{does the rate of increase in the merger rate depend on mass?}

These two questions are not independent.
However, different astrophysical scenarios may be better modeled by one approach over the other (a redshift-dependent mass distribution vs. a mass-dependent redshift distribution).
Different binary formation channels (e.g. stellar clusters, common envelope, stable mass transfer) are generically expected to trace different redshift-dependent merger histories.
If the observed binary black hole population is itself composed of systems from several such channels, each dominating in different mass regimes, then the resulting redshift-dependent merger rate may naturally be well-modeled by Eq.~\eqref{eq:f-z} but with hyperparameters that vary as a function of mass.

To explore this possibility, we follow the same methodology as in Section~\ref{sec:method} but invert the roles of mass and redshift.
Specifically, we elevate the parameters $\alpha_z$, $\beta_z$, and $z_p$ characterizing the redshift-dependent merger rate, see Eq.~\eqref{eq:f-z}, to be functions of primary mass: $\alpha_z(m_1)$, $\beta_z(m_1)$, and $z_p(m_1)$.
Each of these are modelled as sigmoids:
    \begin{equation}
    \label{eq:sigmoid-m}
    \Lambda(m_1) = \frac{\Lambda_{\rm high} - \Lambda_{\rm low}}{1 + \exp\left[-\frac{1}{\Delta m_\Lambda}\left(m_1 - \bar m_\Lambda\right) \right]} + \Lambda_\mathrm{low}.
    \end{equation}
Analogous to Eq.~\eqref{eq: sigmoid delta}, here $\Lambda$ stands for one of the hyperparameters $\{\alpha_z, \beta_z, z_p\}$, with $\Lambda(m_1)$ transitioning from $\Lambda_\mathrm{low}$ at masses $m_1\ll \bar m_\Lambda$ to $\Lambda_\mathrm{high}$ at $m_1 \gg \bar m_\Lambda$ across an interval of scale width $\Delta m_\Lambda$.
Our full model for the redshift-dependent merger rate becomes
    \begin{equation}
    \label{eq:full merger rate alt}
    \begin{aligned}
    &\mathcal{R}(m_1,m_2,\vec\chi_1,\vec\chi_2;z) \\
        &\quad = R_{\rm ref}
            \frac{f(z|m_1)}{f(0.2|20\,M_\odot)}
            \frac{\phi(m_1)}{\phi(20\,M_\odot)}
            p(m_2) p(\vec \chi_1) p(\vec \chi_2);
    \end{aligned}
    \end{equation}
compare to Eq.~\eqref{eq:full merger rate}.

\begin{figure}
    \centering
    \includegraphics{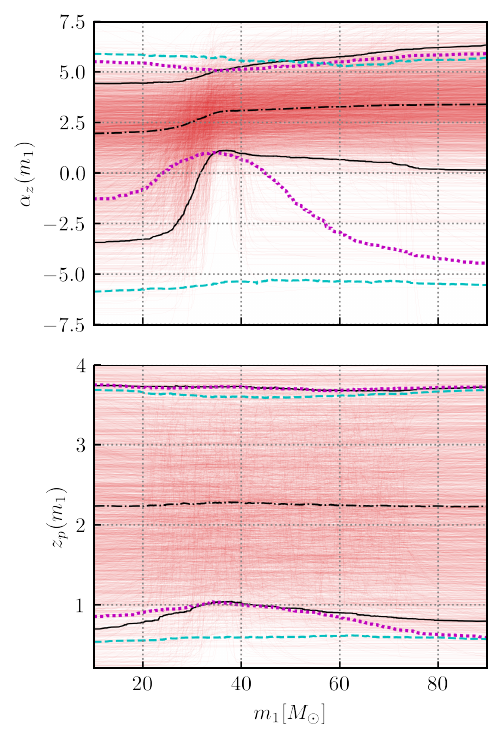}
    \caption{
    Constraints on the power-law slope $\alpha_z$ governing evolution of the volumetric binary black hole merger rate with redshift (\textit{top}) and the peak redshift $z_p$ beyond which the merger rate turns over (\textit{bottom}), each as as function of primary mass.
    As in Figs.~\ref{fig:sigmoid-peak} and \ref{fig:traces-power-law}, black lines give $95\%$ credible posterior bounds, dashed cyan lines mark prior bounds, and dotted magenta curves illustrate prior bounds once conditioned on the ``best-measured'' posteriors at $m_1=35\,M_\odot$.
    Data are consistent with a constant value of $\alpha_z$ (no differential evolution of the black hole merger rate with mass), although the large uncertainties are also consistent with values that vary widely as a function of mass.
    Current data yield mass-dependent lower bounds on $z_p(m_1)$, but otherwise offer no information about possible mass dependence of this peak redshift.}
    \label{fig:sigmoid-mass}
\end{figure}

In Fig.~\ref{fig:sigmoid-mass} we show the resulting posteriors on the low-redshift power-law slope $\alpha_z(m_1)$ and the peak redshift $z_p(m_1)$ as a function of $m_1$.
The data are entirely uninformative regarding the high-redshift slope $\beta_z(m_1)$ and so we do not show a posterior on this parameter.
Our posterior on $\alpha_z(m_1)$ in the upper subplot shows very marked behavior.
At low primary masses $\alpha_z(m_1)$ is broadly constrained to lie between $-4.1\leq \alpha_z \leq 4.3$ at $95\%$ credibility.
The posterior then undergoes a rapid transition around $35\,M_\odot$, such that $\alpha_z(m_1=35\,M_\odot)$ is bounded between $0.7$ and $5.1$ at $95\%$ credibility.
The posterior subsequently slightly broadens again as we look to higher masses.
This behavior is expected.
By virtue of search selection functions, a large number of binaries with $m_1 \approx 35\,M_\odot$ are observed out to large distances, providing a reasonably precise measurement of $\alpha_z$ in this mass range.
In contrast, few low-mass binaries are observed, with those that are detected occuring at small redshifts.
Thus we currently have little information about $\alpha_z$ for low-mass events.
The fact that we detect a non-zero number of massive binaries at large redshifts also yields a lower limit on $\alpha_z$ for high-mass events, although with somewhat broader uncertainties set by the relatively small number of such events.

Analogously to Figs.~\ref{fig:sigmoid-peak} and \ref{fig:traces-power-law}, the dotted magneta curves show conditional priors on $\alpha_z(m_1)$.
Note that, in these previous figures, we showed priors conditioned on measurements at $z=0$, this being the redshift at which we obtained the most precise measurement of the black hole mass distribution.
In the present context, it is clear that the black hole redshift distribution is most precisely measured at $m_1\approx 35\,M_\odot$.
We accordingly condition priors at this location, yielding the ``waist'' appearing in the upper panel of Fig.~\ref{fig:sigmoid-mass}.
The deviation between our $\alpha_z(m_1)$ posterior and the conditional prior indicates the presence of additional information in the data.
At high masses, we see that data disfavor small or negative values of $\alpha_z$.
At low masses, although the posterior and conditioned prior largely coincide, the posterior is nevertheless pushed slightly to smaller values of $\alpha_z$.
This may hint at distinct redshift evolution of low- and high-mass binary black holes.

The lower subplot of Fig.~\ref{fig:sigmoid-mass} similarly shows our prior, posterior, and conditional prior on $z_p(m_1)$.
The peak redshift is bounded away from zero in the $30$--$50\,M_\odot$ range, due to the large number of events observed in this mass range.
Otherwise, the posterior is largely uninformative, closely following the conditional prior at lower and higher masses.

Despite the non-trivial constraints on $\alpha_z(m_1)$, on the whole we find no requirement that the black hole redshift distribution varies as a function of mass.
Current data remain consistent with a population whose merger rate evolves synchronously across the full mass range, with single, universal values of $\alpha_z$ and $z_p$.
Additional results are discussed in Appendix~\ref{app: inference results}, including corner plots illustrating full posteriors on the hyperparameters governing $\alpha_z(m_1)$ and $z_p(m_1)$.

\section{Conclusion}
\label{sec:conclusion}

In this paper we have systematically surveyed current gravitational-wave data for redshift-dependence in the binary black hole mass spectrum.
We found no evidence for a correlation between black hole masses and redshifts, and present-day data are consistent with a primary mass spectrum that remains stationary with redshift.

In some cases, we identify non-trivial constraints on the degree of redshift evolution allowed by current data.
We found in Sec.~\ref{sec:peak-sector} that the location of the $35\,M_\odot$ peak in the black hole primary mass spectrum is measured to remain relatively fixed across a broad range of redshifts.
Similarly, in Sec.~\ref{sec:power-law-sector} we found that the slope of the power-law continuum is approximately constant out to redshift $z\approx 1$.
These constraints may already be sufficient to rule out theoretical models that predict largescale evolution of black hole masses with redshift.
At the same time, systematic variation of black hole masses with redshift is not excluded.
For example, current data are still consistent with scenarios in which the height of the $35\,M_\odot$ peak rises or falls somewhat considerably with redshift (see Sec.~\ref{sec:results}).
We emphasize, however, that such evolution is \textit{not a requirement of the data}.
This is in tension with the conclusions drawn in~\cite{Karathanasis_2023} and~\cite{Rinaldi}, but is consistent with results of other analyses~\citep{Fishbach_2021,van_Son_2022,PhysRevX.11.021053,2023ApJ...957...37R,2024arXiv240616844H}.
While our work was reaching its final stages of preparation, we were additionally made aware of a complementary study applying the methodology of~\cite{2024ApJ...960...65S} to explore redshift variation of black hole masses~\citep{Sadiq_inPrep}; this study finds no statistically significant evidence for an evolving mass distribution with GWTC-3, consistent with our results.

Although current data do not yet indicate a correlation between black hole redshifts and mass, it would be surprising if no correlation was fundamentally present, given the particularly large number of ways such correlations may arise (see Sec.~\ref{sec:intro}).
Instrumental upgrades and continued commissioning leading up to and during the current LIGO-Virgo-KAGRA O4 observing run increase the distance to which binary black holes can be detected and the precision with which they can be characterized~\citep{capote_advanced_2024}, and we anticipate that future catalogs may yet reveal a black hole mass spectrum that changes over cosmic time.
If a mass-redshift correlation remains unobserved, however, the astrophysical implications of a null result become increasingly impactful.
A mass distribution that is known to remain constant over a wide range of redshifts may indicate that binary black holes experience very long time delays between their formation and eventual merger, ``washing out'' the time varying conditions present at their birth.
It could indicate that massive black hole formation is much less dependent on stellar metallicity than typically expected, and/or that metal-poor environments remain prevalent even at late cosmic times.
Finally, it would likely indicate that no more than one formation channel contributes non-negligibly to the observed population of black hole mergers.\\ \\

K.~T.~was supported by FWO-Vlaanderen through grant number 1179524N. 
T.~C.~is supported by the Eric and Wendy Schmidt AI in Science Postdoctoral Fellowship, a Schmidt Sciences program.
This research has made use of data, software and/or web tools obtained from the Gravitational Wave Open Science Center (\url{https://www.gw-openscience.org}), a service of LIGO Laboratory, the LIGO Scientific Collaboration and the Virgo Collaboration. 
Virgo is funded by the French Centre National de Recherche Scientifique (CNRS), the Italian Istituto Nazionale della Fisica Nucleare (INFN) and the Dutch Nikhef, with contributions by Polish and Hungarian institutes. 
This material is based upon work supported by NSF’s LIGO Laboratory which is a major facility fully funded by the National Science Foundation. 
The authors are grateful for computational resources provided by the LIGO Laboratory and supported by NSF Grants PHY-0757058 and PHY-0823459.\\ \\

\textit{\large Data \& code availability:}
Code used to perform this study is made available on GitHub at \textcolor{blue}{\url{https://github.com/maxlalleman/bbh_mass_distribution_redshift_variation_inference}}, and datasets comprising our results are available on Zenodo at \textcolor{blue}{\url{https://zenodo.org/records/14671139}}.

\software{arviz~\citep{arviz_2019}, 
          astropy~\citep{2013A&A...558A..33A,2018AJ....156..123A}, 
          jax~\citep{jax2018github}, 
          matplotlib~\citep{Hunter:2007}, 
          numpy~\citep{harris2020array}, 
          numpyro~\citep{phan2019composable, bingham2019pyro}, 
          scipy~\citep{2020SciPy-NMeth}
          }

\appendix
\renewcommand{\thefigure}{A\arabic{figure}}
\renewcommand{\thetable}{A\arabic{table}}
\setcounter{figure}{0}
\setcounter{table}{0}

\section{Spin Models and Prior Information}
\label{app: prior}

In this appendix we provide more details regarding our spin models (not discussed in the main text) and the priors adopted on our hyperparameters during our population inference.

We assume that dimensionless component spin magnitudes are independently and identically modeled as truncated Gaussians:
\begin{equation}
    p(\chi_i)=\sqrt{\frac{2}{\pi\sigma_\chi^2}}\frac{e^{-(\chi_i-\mu_\chi)^2/2\sigma_\chi^2}}{{\rm Erf}\left(\frac{1-\mu_\chi}{\sqrt{2\sigma_\chi^2}}\right)+{\rm Erf}\left(\frac{\mu_\chi}{\sqrt{2\sigma_\chi^2}}\right)},
\end{equation}
with mean $\mu_\chi$ and variance $\sigma_\chi^2$.
Similarly, we assume that cosine spin-orbit misalignment angles are distributed as truncated Gaussians centered at $\cos\theta = 1$:
\begin{equation}
    \pi(\cos\theta_i)=\sqrt{\frac{2}{\pi\sigma_{u}^2}}\frac{e^{-(\cos\theta_i-1)^2/2\sigma_{u}^2}}{{\rm Erf}\left(\frac{-2}{\sqrt{2\sigma_{u}^2}}\right)},
\end{equation}
with variance $\sigma_{u}^2$ to be inferred from the data.

Table \ref{tab:prior tabel}, meanwhile, gives the priors used in our analysis.
When exploring redshift variation in a given mass hyperparameter $\Lambda$, identical priors $p(\Lambda_\mathrm{high})$ and $p(\Lambda_\mathrm{low})$ are placed on the asymptotic high- and low-redshift values.
The same is true when conversely exploring mass variation in hyperparameters governing the black hole redshift distribution.
We note that, when transition redshifts $\overline z_\Lambda$ were allowed to extend beyond $\overline z_\Lambda = 0.8$, we encountered extreme sampling difficulties related to poorly-converged Monte Carlo averages.
We therefore impose a prior that limits $\overline z_\Lambda\leq 0.8$.

\begin{table}[!h]
    \caption{
    Priors adopted on the hyperparameters used in our analyses.
    $\mathcal{U}(a,b)$ indicates a uniform prior normalized between $a$ and $b$, and $\mathcal{N}(a,b)$ a Gaussian prior with mean $a$ and standard deviation $b$.
    When varying a given hyperparameter as a function of mass or redshift, priors on the asymptotic values are identical to the priors listed in the upper portion of the table.
    When varying the mean of the Gaussian peak in the black hole mass spectrum, for example, we adopt $p(\mu_{m,\mathrm{low}}) = p(\mu_{m,\mathrm{high}}) = p(\mu_m) = \mathcal{U}(15\,M_\odot,60\,M_\odot)$.
    }
    \vspace{0.3cm}
    \centering
    \begin{tabular}{r|l}
        \hline
        \hline
         Hyperparameter & Prior \\
         \hline
         $\log(\mathcal{R}_{\rm ref}\cdot\mathrm{Gpc}^{3}\,\mathrm{yr})$ & $\mathcal{U}$(-2, 1) \\
         $\alpha$ & $\mathcal{N}(-2, 3)$ \\
         $\log(f_{\rm p}$) & $\mathcal{U}$(-6, 0) \\
         $\beta_q$ & $\mathcal{N}$(0, 3) \\
         $\alpha_z$ & $\mathcal{N}$(0, 4) \\
         $\beta$ & $\mathcal{U}$(0, 10) \\
         $z_p$ & $\mathcal{U}$(0.2, 4) \\
         $M_{\rm min}$ & $\mathcal{U}(5\,M_\odot, 15\,M_\odot)$ \\
         $M_{\rm max}$ & $\mathcal{U}(50\,M_\odot, 100\,M_\odot)$ \\
         $\mu_{m}$ & $\mathcal{U}(15\,M_\odot, 60\,M_\odot)$ \\
         $ \sigma_{m}$ & $\mathcal{U}(1.5\,M_\odot, 15\,M_\odot)$ \\
         $\log(\delta m_{\rm min}/\,M_\odot)$ & $\mathcal{U}(-1, 0.5)$ \\
         $\log(\delta m_{\rm max}/\,M_\odot)$ & $\mathcal{U}(0.5, 1.5)$ \\
         \hline
         $\Lambda_\mathrm{high}$ \& $\Lambda_\mathrm{low}$ & Identical prior to $p(\Lambda)$ above \\
         $\log(\Delta z_\Lambda)$ & $\mathcal{U}$(-1, 1) \\
         $\bar z_{\Lambda}$ & $\mathcal{U}(0, 0.8)$ \\
         $\log(\Delta m_\Lambda/M_\odot)$ & $\mathcal{U}$(-1, 3) \\
         $\bar m_{\Lambda}$ & $\mathcal{U}(20\,M_\odot, 75\,M_\odot)$ \\
         \hline
         \hline
    \end{tabular}
    \label{tab:prior tabel}
\end{table}

\section{Data}\label{app: data}
\begin{figure*}[t]
	\centering
	\includegraphics[width=0.9\textwidth]{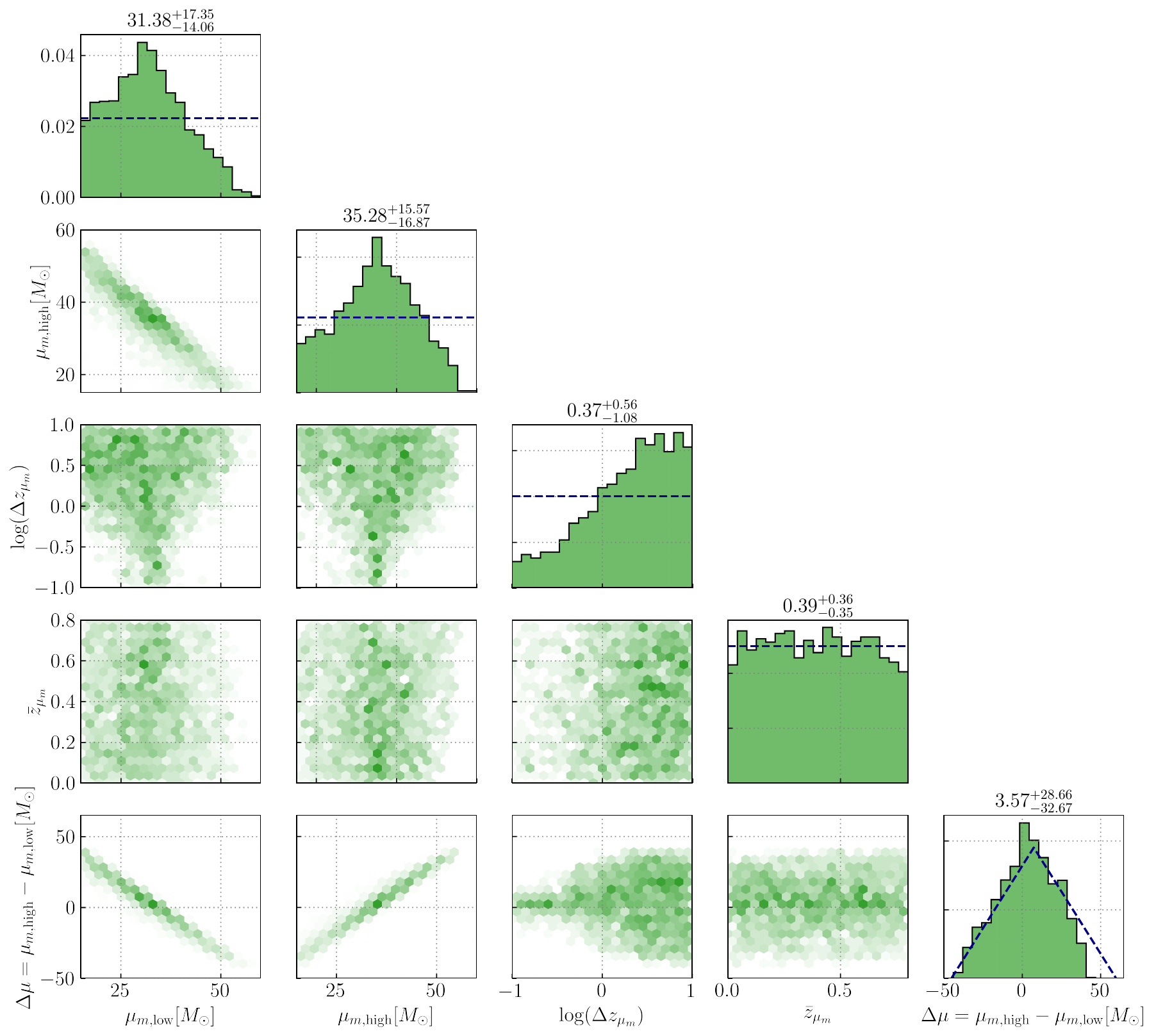}
	\caption{Posteriors on the hyperparameters controlling the variation of $ \mu_{m}$ with redshift.
		Specifically, we show the asymptotic low-redshift peak location $\mu_{m,\mathrm{low}}$, the asymptotic high-redshift value $\mu_{m,\mathrm{high}}$, the logarithmic scale $\log(\Delta z_{\mu_m})$ over which the transition occurs, the transition location $\overline z_{\mu_m}$, and the derived posterior on the difference $\Delta \mu = \mu_{m,\mathrm{high}}- \mu_{m,\mathrm{low}}$.
		The blue dashed lines in the one-dimensional posterior give the priors placed on each parameter.
		The difference $\Delta \mu$ between high- and low-redshift peak locations is consistent with zero, indicating no evidence for redshift-dependence.
		Note also that the posterior on $\Delta z_{\mu_m}$ is skewed towards higher values, with large scale lengths required when $\Delta \mu$ is non-zero in order to guarantee a non-evolving peak location in the redshift range of interest.}
	\label{fig:PE-sigma}
\end{figure*}

This appendix details the exact datasets used in our analysis.
We take as inputs the binary black holes contained in the GWTC-3 catalog~\citep{PhysRevX.13.011048} released by the LIGO-Virgo-KAGRA Collaboration.
In particular, we select all binaries detected with false alarm rates below one per year, except for two events (GW190814 and GW190917) that are known to be population outliers~\citep{Abbott_2020,PhysRevX.13.011048}; these are excluded from our analysis.
Parameter estimation samples for each binary black hole are accessed through the Gravitational-Wave Open Science Center\footnote{https://www.gw-openscience.org/} \citep{Vallisneri_2015,Rich_Abbott_2021,KAGRA:2023pio} or via Zenodo\footnote{https://zenodo.org/record/5546663}.
For events first published in GWTC-1~\citep{PhysRevX.9.031040}, we use the ``Overall posterior'' samples.
For events first published in GWTC-2~\citep{PhysRevX.11.021053}, we adopt the ``PrecessingSpinIMRHM'' samples, while we use the ``C01:Mixed'' samples\footnote{https://zenodo.org/record/5546663} for events first published in GWTC-3~\citep{PhysRevX.13.011048}.

As described in the main text, selection effects are calculated using Monte Carlo averages over sets of mock signals injected into and recovered from gravitational-wave data.
We adopt the injection sets described in \cite{PhysRevX.13.011048}.
For injections performed in the O3 observing run, we consider them detected if they are recovered with a false-alarm rate below one per year in at least one search pipeline, matching our event selection criteria above.
Injections performed in O1 and O2 only have network signal-to-noise ratios, not false alarm rates; for these events we demand that detected events have signal-to-noise ratios above $10$.

\section{Additional inference results}\label{app: inference results}

In this appendix we present and discuss additional inference results from our analyses in Secs.~\ref{sec:results} and \ref{sec: reverse sector}, including corner plots showing more detailed posteriors on our population parameters.

Figure~\ref{fig:PE-sigma} shows our posterior on the hyperparameters governing redshift evolution of the Gaussian peak in the binary black hole primary mass spectrum: the asymptotic low-redshift value $\mu_{m, \rm low}$, the high-redshift value $\mu_{m, \mathrm{high}}$, the transition redshift $\bar z_{\mu_m}$, and the scale $\Delta z_{\mu_m}$ over which the transition occurs.
\begin{figure*}[t]
	\centering
	\includegraphics[width=0.9\textwidth]{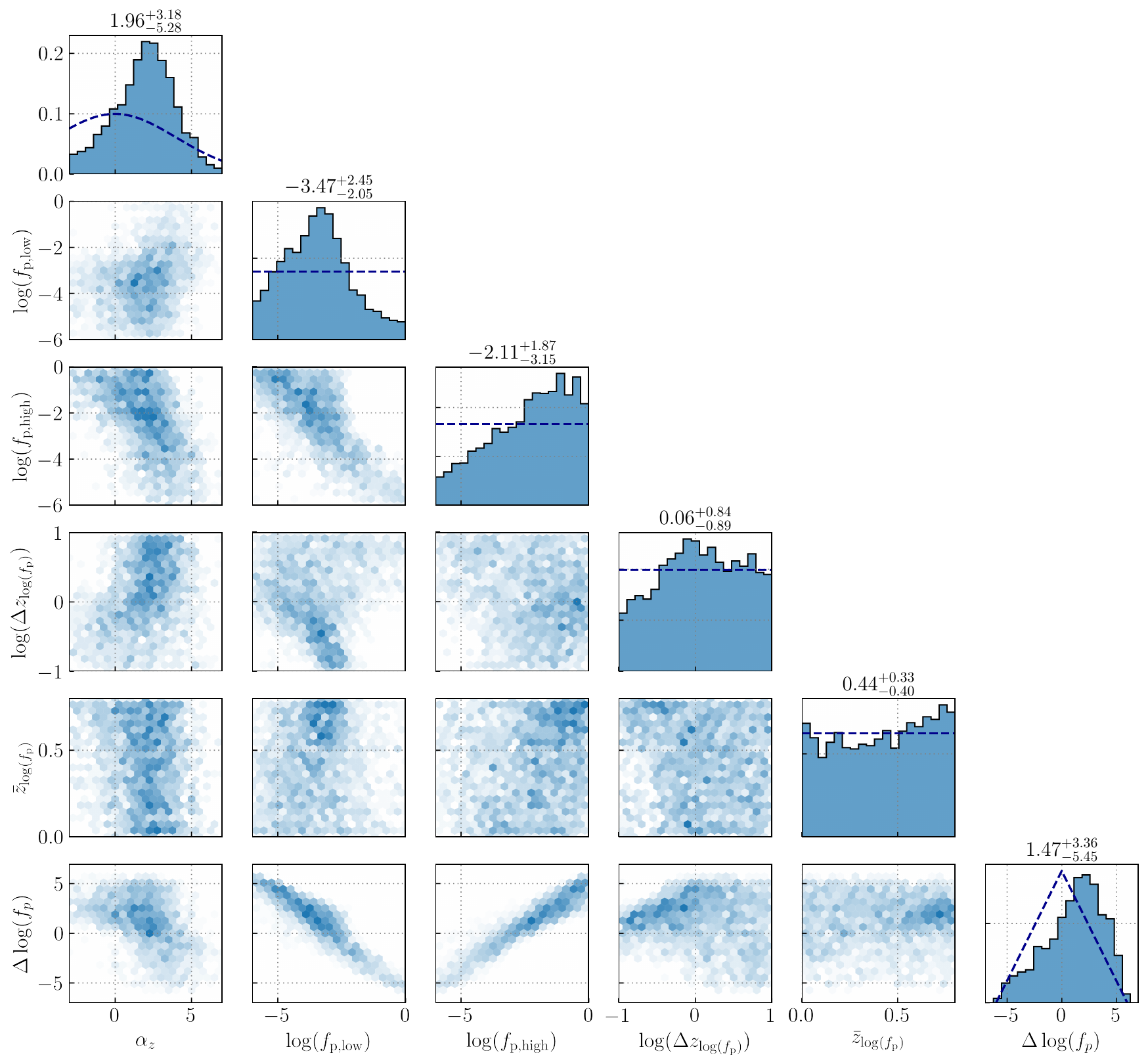}
	\caption{Various one-dimensional and two-dimensional posteriors of the hyperparameters in our analysis relating to the variation of $f_{\text{p}}$ in the power-law continuum are shown. The dark blue dashed lines in the one-dimensional distributions represent the priors. The hyperparameters shown here are the (ascending) power-law index $\alpha_z$ of the modeled merger rate, and hyperparameters connected to varying $f_{\rm p}$ from Eq. \eqref{eq: sigmoid delta}. This includes $\log(f_{\rm p, low})$, $\log(f_{\rm p, high})$, the width of the sigmoid $\log(\Delta z_{\log(f_{\rm p})})$ and the middle of the sigmoid $\bar z_{\log(f_{\rm p})}$. We observe a slight increase in the posterior for $\bar z_{\log(f_{\rm p})}$ values around 0.5-0.8. Variation in redshift of $f_\mathrm{p}$ is allowed, but not discovered: the posterior on $\log(f_{p, \rm high})$ is broad, but with a slight preference towards more positive and therefore greater values. Smaller values are also disfavored for the width of the sigmoid.}
	\label{fig:PE_power-law-only-fpeak}
\end{figure*}
For convenience, we also present the derived posterior on the difference $\Delta  \mu_{m} = \mu_{m,\mathrm{high}} -  \mu_{m,\mathrm{low}}$ between the asymptotic peak locations.
We see that, although redshift variation in the peak location is not excluded, a non-zero value of $\Delta  \mu = \mu_{\rm high} -  \mu_{m}$ is not required, consistent with a non-evolving peak.
Note that $|\Delta \mu|$ is permitted to be large only when the scale length $\Delta z_{\mu_m}$ of the transition is also large, such that the peak location is still forced to remain relatively constant over the range of observable redshifts.
This same behavior also manifests as a strong anticorrelation between $\mu_{m,\mathrm{low}}$ and $\mu_{m,\mathrm{high}}$, which together must conspire to preserve $\mu_m \approx 35\,M_\odot$ across the range of observable redshifts (see Fig.~\ref{fig:sigmoid-peak}).

Similarly, Fig.~\ref{fig:PE_power-law-only-fpeak} shows our posterior on hyperparameters connected to redshift-variation of the height $f_p$ of the Gaussian peak.
We specifically show the posterior derived in Sec.~\ref{sec:power-law-sector}, although these results are extremely similar to the analogous constraints on $f_p(z)$ derived in Sec.~\ref{sec:peak-sector}, as well as in Appendix~\ref{app:varying-all} when varying all features in the black hole mass spectrum simultaneously.
As discussed in the main text, the net change $\Delta \log(f_p) = f_{\rm p, high} - f_{\rm p, low}$ in the peak height with redshift is consistent with zero, although with a preference for positive values (a rising peak) over negative values (a shrinking peak).
\begin{figure*}
	\centering
	\includegraphics[width=0.9\textwidth]{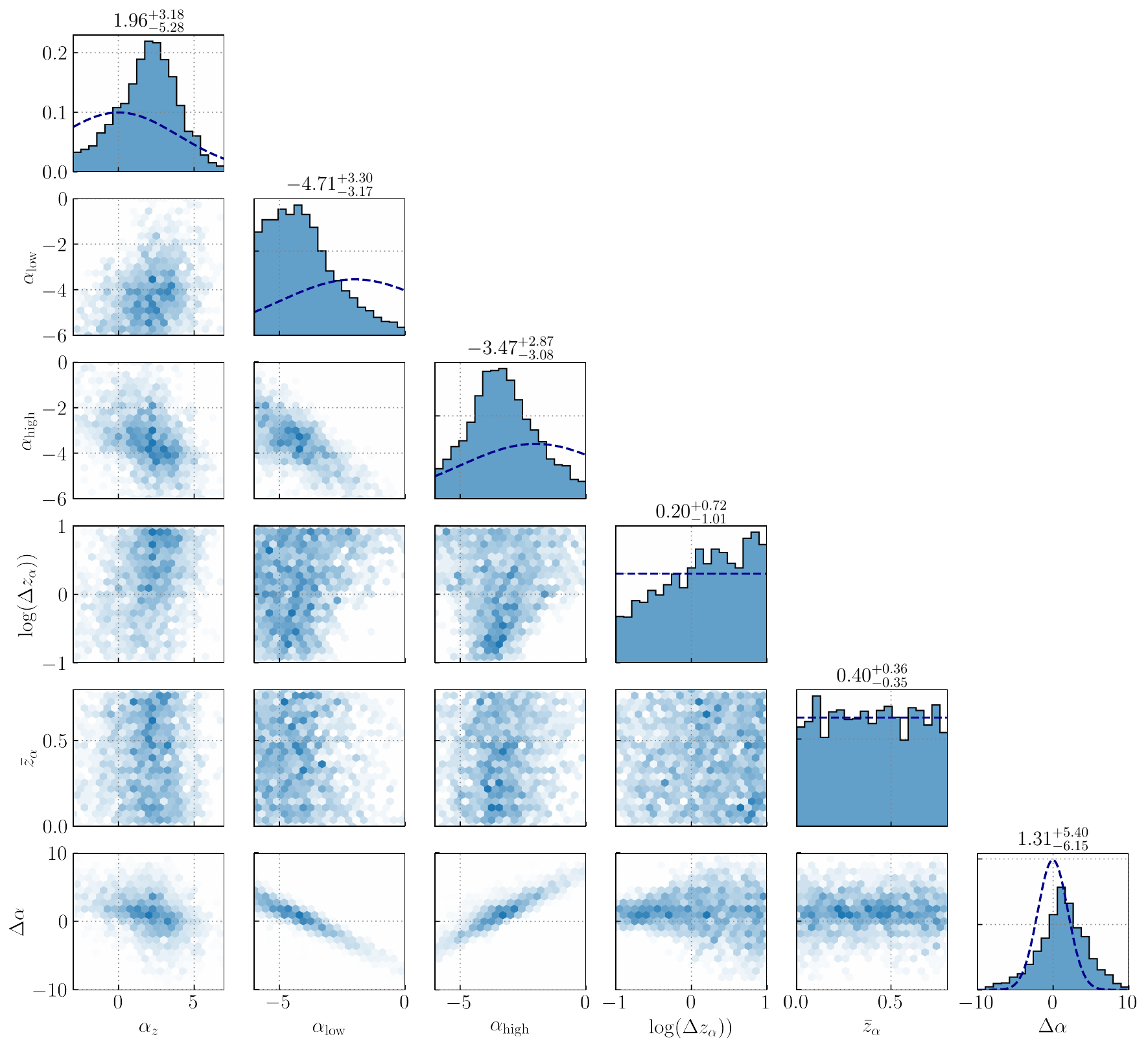}
	\caption{Posteriors of the hyperparameters related to the variation of $\alpha$ in redshift are shown. The dark blue dashed lines in the one-dimensional distributions represent the priors. The hyperparameters shown here are the (ascending) power-law index $\alpha_z$ of the modeled merger rate, and then hyperparameters connected to varying $\alpha$ from Eq. \eqref{eq: sigmoid delta}. This includes $\alpha_{\rm low}$, $\alpha_{\rm high}$, the width of the sigmoid $\log(\Delta z_{\alpha})$ and the middle of the sigmoid $\bar z_{\alpha}$. We also plot the difference between the high-redshift and low-redshift value $\Delta \alpha$, which now is a Gaussian, because the difference of two Gaussians is once again a Gaussian distribution variable. We see a preference for larger values of the sigmoid width $\log(\Delta z_{\alpha})$ and larger values of the difference $\Delta \alpha$.}
	\label{fig: PE-power-law-alpha}
\end{figure*}
As was the case with $\mu_m(z)$ in Fig.~\ref{fig:PE-sigma}, any large changes in the peak height are generally required to occur over long scale lengths $\Delta z_{\log f_p}$
We also include in Fig.~\ref{fig:PE_power-law-only-fpeak} the posterior on the slope $\alpha_z$ with which the overall merger rate increases with redshift.
This parameter is somewhat anticorrelated with $\Delta \log(f_p) = f_{\rm p, high} - f_{\rm p, low}$.
This is expected: in order to correctly predict the observed number of $35\,M_\odot$ events at large redshifts without \textit{overpredicting} the number of such events at small redshifts, the model can either posit that the overall merger rate grows with redshift (large $\alpha_z$ and small $\Delta f_p$), or adopt a constant merger rate but invoke a growing peak (small $\alpha$ and large $\Delta f_p$).

\begin{figure*}
    \centering
    \includegraphics[width=0.9\textwidth]{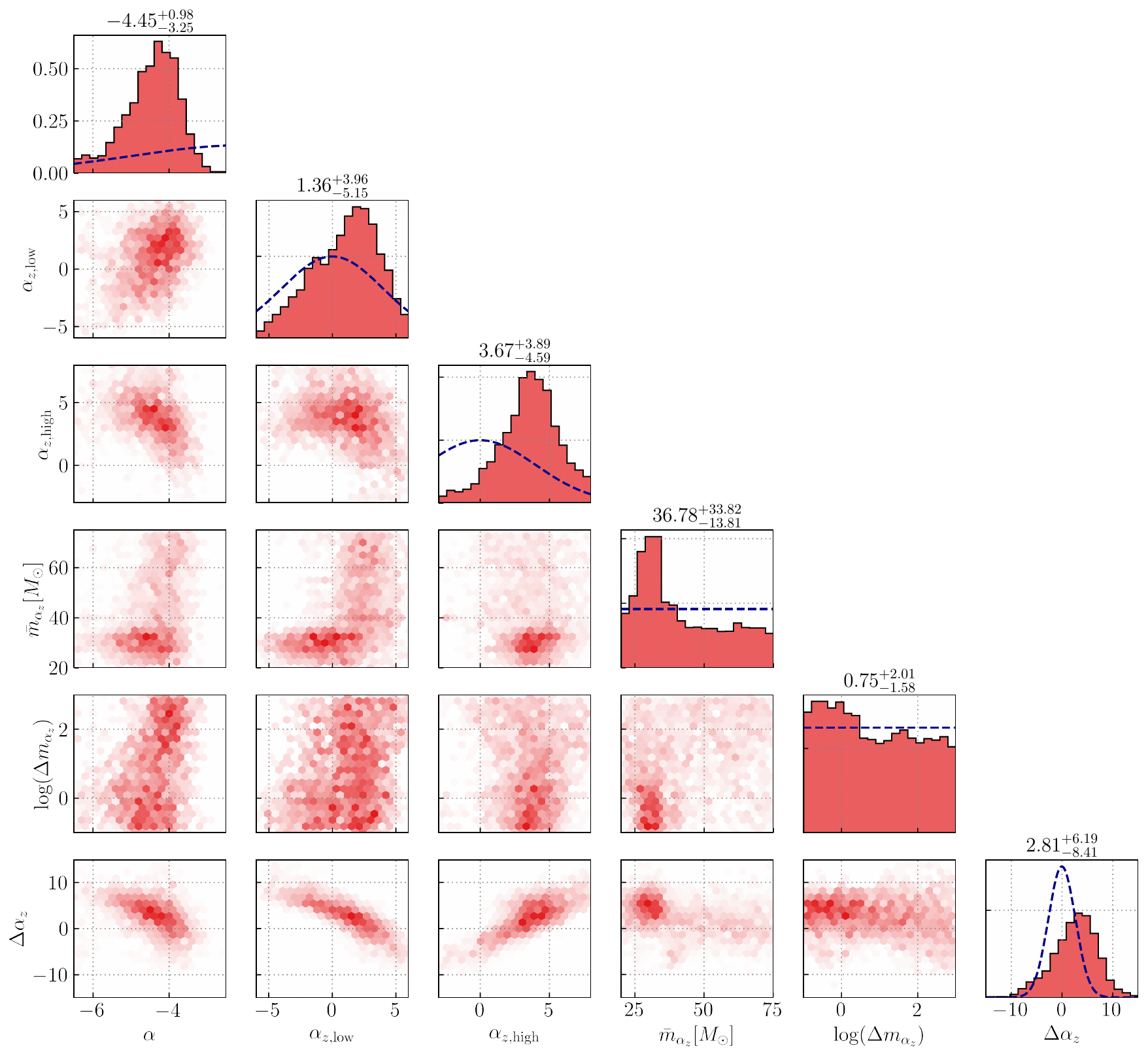}
    \caption{Posterior plots for varying the hyperparameters connected to $\alpha_z$. In this figure the power-law index of the primary mass distribution $\alpha$, the four hyperparameters connected to the variation of $\alpha_z$ via Eq. \eqref{eq: sigmoid delta}, and the posterior difference $\Delta \alpha_z$, are considered. A clear peak is visible at 33 $M_\odot$ for the middle point of the sigmoid $\bar{m}_{\alpha_z}$, coinciding with the view provided by the posterior samples in Figure \ref{fig:sigmoid-mass}. We also observe a preference for higher values of $\alpha_{z, \rm high}$ compared to $\alpha_{z, \rm low}$. That is confirmed by considering the posterior difference that shows a preference for positive values, but not an exclusion of zero and hence no variation.}
    \label{fig:pe-alpha-small}
\end{figure*}

The same broad features are present in Fig.~\ref{fig: PE-power-law-alpha}, which presents the posterior controlling the power-law slope of the primary mass spectrum.
We again see a slight negative correlation between $\alpha_{\rm low}$ and the ascending power-law index $\alpha_z$ controlling growth of the overall merger rate.
Large $\alpha_z$ will generally overpredict the number of massive events at large redshift, unless compensated for by small $\Delta \alpha$.
Conversely, large $\Delta \alpha$ must in turn be compensated by small $\alpha_z$ to guarantee less frequent massive mergers.

\begin{figure*}
    \centering
    \includegraphics[width=0.9\textwidth]{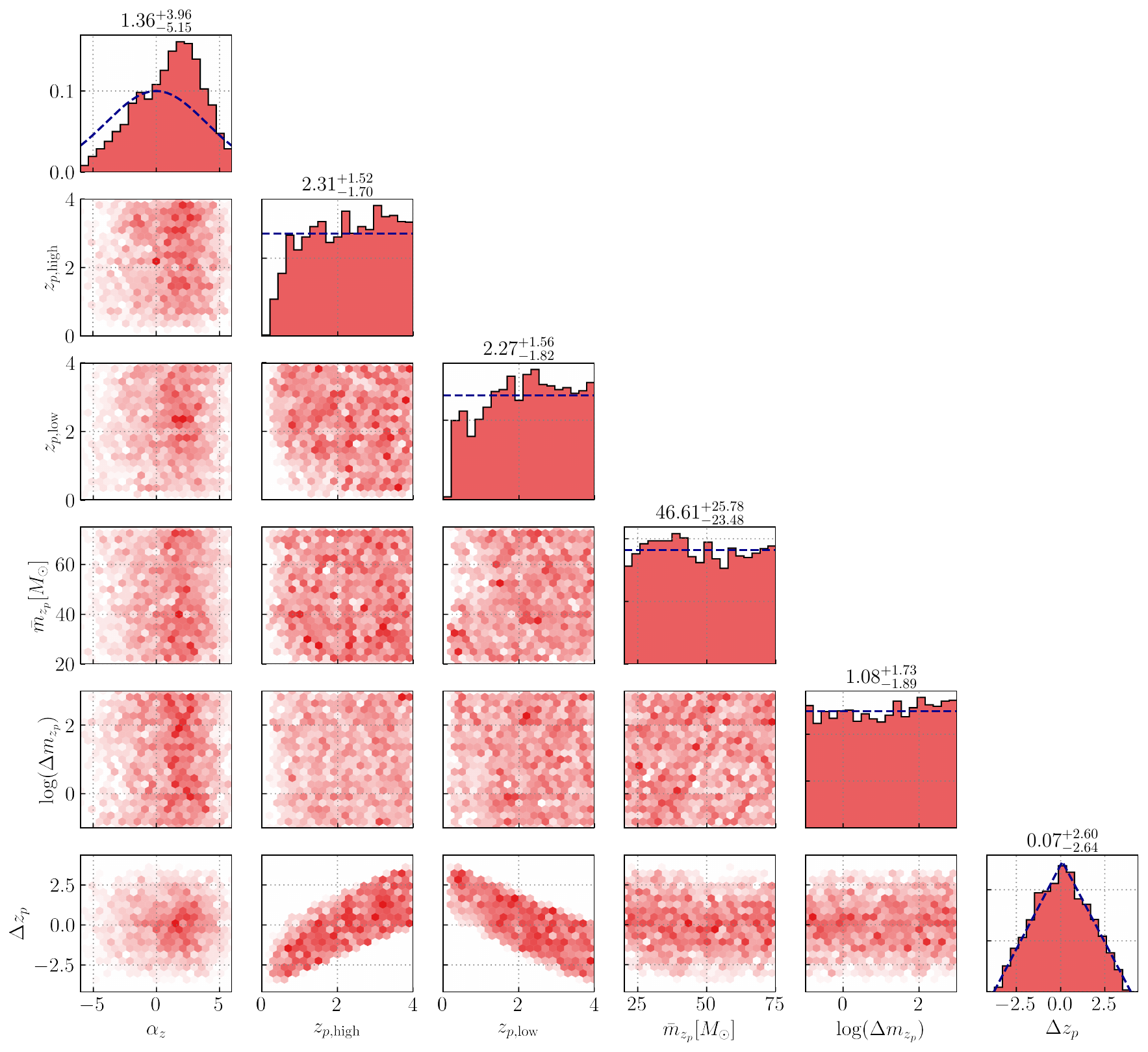}
    \caption{Posterior plots for all hyperparameters connected to the variation of $z_p$ in $m_1$ are shown. The dashed lines denote the one-dimensional priors in the one-dimensional plots on top. The two-dimensional posteriors are shown in the corresponding row and column for the hyperparameters. Hyperparameters in this figure include the ascending power-law index $\alpha_z$, the four hyperparameters connected to varying $z_p$ via Eq. \eqref{eq: sigmoid delta}, and the difference between the high- and low-redshift value of $z_p$. No significant variation is identified, however there is some preference for higher values of $z_{p, \mathrm{high}}$.}
    \label{fig:Pe-zp}
\end{figure*}

\begin{figure}
	\includegraphics{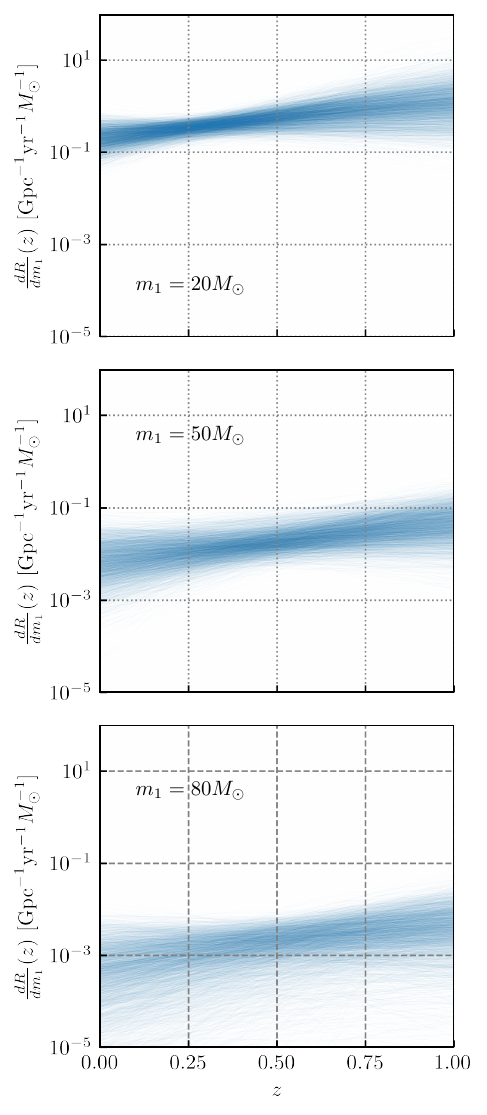}
	\caption{Three slices of the differential merger rate consisting out of the power-law analysis samples but viewed at a different $m_1$ are shown. Above, we show $m_1$ = 20 $M_\odot$, in the middle $m_1$ = 50 $M_\odot$ and below $m_1$ = 80 $M_\odot$. For higher masses, the merger rate goes down, in accordance with the primary mass distribution, while the merger rate is also more spread out at these masses. A broadening of the samples when moving from redshift $z$ = 0.5 to $z$ = 1 is also observed.}
	\label{fig: merger-rate-for-different-m}
\end{figure}

An alternative view of our results from Sec.~\ref{sec:power-law-sector} is given in Fig.~\ref{fig: merger-rate-for-different-m}.
In this figure, we show our posterior on the redshift-dependent merger rate (Eq.~\eqref{eq:full merger rate alt}) evaluated at three different primary mass values.
The absolute merger rates in each panel reflect the overall shape of the mass distribution, which favors low-mass mergers, and uncertainties increase towards both larger masses and larger redshifts.
At the same time, the merger rates in at each of the three masses are seen to rise in unison with one another, with no mass-dependent deviations.

In Sec.~\ref{sec: reverse sector} we reversed our approach and instead investigated \textit{mass-dependence} in the binary black hole redshift distribution.
Figures~\ref{fig:pe-alpha-small} and \ref{fig:Pe-zp} show posteriors on parameters governing mass evolution of the low-redshift slope $\alpha_z$ of the merger rate and the peak redshift $z_p$, respectively.
Figure~\ref{fig:pe-alpha-small} exhibits a significant amount of structure.
Exactly as discussed above, we see an anticorrelation between the slope $\alpha$ of the black hole primary mass spectrum and any shifts $\Delta \alpha_z$ in the growth of the merger rate as a function of mass.
As illustrated in Fig.~\ref{fig:sigmoid-mass}, the data also marginally favor positive $\Delta \alpha_z$, although $\Delta \alpha_z = 0$ remains quite consistent with observation.
The posterior on the transition point $\overline m_{\alpha_z}$, meanwhile, exhibits a somewhat striking peak near $33\,M_\odot$ (this behavior too can be seen in Fig.~\ref{fig:sigmoid-mass}); if there \textit{exists} a transition in $\alpha_z$, it is likely to be centered at this location.
This behavior, however, could simply be a consequence of the fact that a large number of observations occur near this mass, making this the point where $\alpha_z(m_1)$ is best constrained.
Whether this posterior feature is real or just due to the relative underabundance of higher- and lower-mass events will require more data to determine.
Figure~\ref{fig:Pe-zp}, in contrast, is relatively featureless, other than a preference against very low values of $z_{p, \mathrm{low}}$ and $z_{p, \mathrm{high}}$.

\begin{figure}
    \centering
    \includegraphics{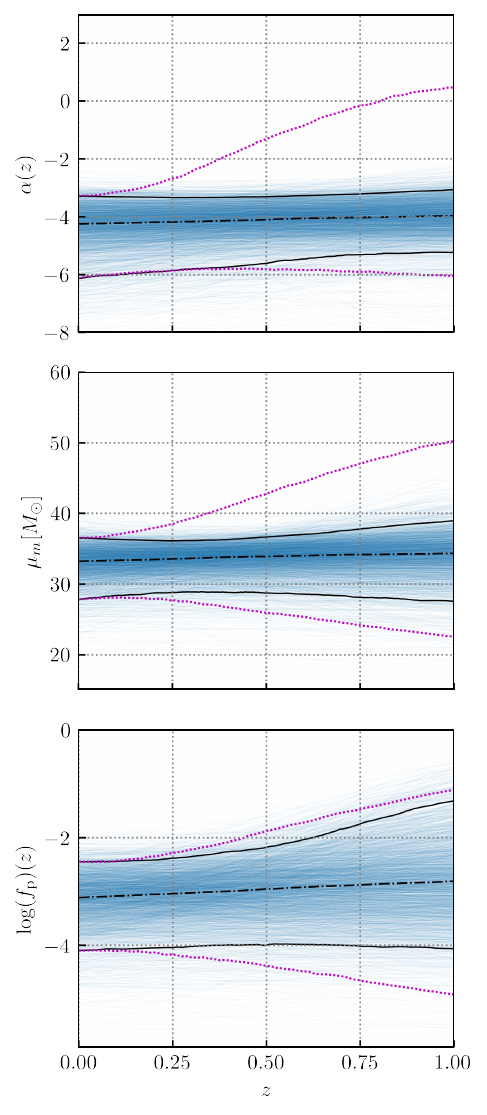}
    \caption{The traces of three hyperparameters from the all-varying analysis, where the blue lines are the sample traces, the dot-dashed black line is the median value of the traces, the solid black lines are the 95$\%$ confidence intervals and the magenta dotted lines are the conditional priors. The top subplot shows the power-law continuum index $\alpha$, the middle subplot the location of the Gaussian excess $\mu_m$ and the bottom subplot shows the fraction of events in the peak $\log f_p$.}
    \label{fig:traces-all_varied}
\end{figure}

\section{Simultaneously varying the Gaussian peak and power-law continuum}
\label{app:varying-all}
\begin{figure}
	\centering
	\includegraphics[width=0.9\linewidth]{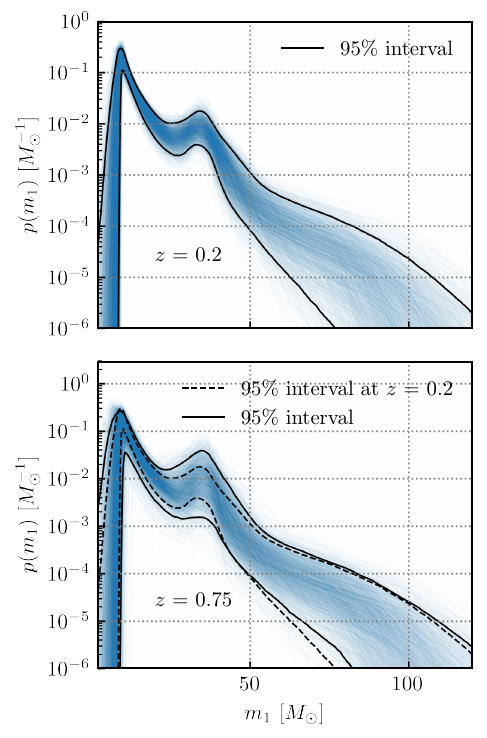}
	\caption{The primary mass distribution is shown for two different redshifts: $z$ = 0.2 and $z$ = 0.75, where the solid black lines show the 95$\%$ confidence intervals for the traces at that redshift, while the dashed black lines in the lower subplot show the same lines but for $z$ = 0.2 for visual comparison.}
	\label{fig:all-varied-p(m)}
\end{figure}

In Secs.~\ref{sec:peak-sector} and \ref{sec:power-law-sector} we independently allowed the Gaussian peak and power-law components of the black hole primary mass spectrum to vary with redshift.
For completeness, we have also repeated our analysis but simultaneously allowing \textit{all} hyperparameters governing the black hole mass distribution to evolve with redshift.

When simultaneously varying both the Gaussian peak and the power-law continuum in this manner, results are unchanged relative to those presented in Sec.~\ref{sec:results}.
In Fig.~\ref{fig:traces-all_varied}, we show posteriors on the three most-informed parameters, the power-law slope and the Gaussian's mean and height, as a function of redshift.
These posteriors exhibit the same trends identified in the main text above (compare with Figs.~\ref{fig:sigmoid-peak} and \ref{fig:traces-power-law}), with no requirement that any systematically vary with redshift.
We also show in Fig.~\ref{fig:all-varied-p(m)} the corresponding posteriors on the primary mass distribution, as measured at redshifts $z=0.2$ and $z=1$; compare with Figs.~\ref{fig:peak-R-m} and \ref{fig: merger-rate-for-different-z}, in which the Gaussian peak and power-law continuum are varied separately.


\bibliography{sample631}{}
\bibliographystyle{aasjournal}

\end{document}